\documentclass{vgtc}                          % final (conference style)
\ifpdf%                                % if we use pdflatex
  \pdfoutput=1\relax                   % create PDFs from pdfLaTeX
  \usepackage{graphicx}                % allow us to embed graphics files
  \DeclareGraphicsExtensions{.pdf,.png,.jpg,.jpeg} % for pdflatex we expect .pdf, .png, or .jpg files
\else%                                 % else we use pure latex
  \usepackage{graphicx}                % allow us to embed graphics files
  \DeclareGraphicsExtensions{.eps}     % for pure latex we expect eps files
\fi%

\graphicspath{{figures/}{pictures/}{images/}{./}} % where to search for the images

\usepackage{microtype}                 % use micro-typography (slightly more compact, better to read)
\PassOptionsToPackage{warn}{textcomp}  % to address font issues with \textrightarrow
\usepackage{textcomp}                  % use better special symbols
\usepackage{mathptmx}                  % use matching math font
\usepackage{times}                     % we use Times as the main font
\usepackage{cite}                      % needed to automatically sort the references
\usepackage{tabu}                      % only used for the table example
\usepackage{booktabs}                  % only used for the table example
\usepackage{multirow}
\usepackage{colortbl}
\usepackage{graphicx}
 \usepackage{float}
 \usepackage[dvipsnames,table,xcdraw]{xcolor}
 \usepackage{amsfonts}
 \usepackage[T1]{fontenc}
\usepackage{soul}
\usepackage[inline]{trackchanges}

\onlineid{1128}

\vgtccategory{Research}

\vgtcinsertpkg

\title{VRMN-bD: A Multi-modal Natural Behavior Dataset of Immersive Human Fear Responses in VR Stand-up Interactive Games}

\author{He Zhang\thanks{e-mail: hpz5211@psu.edu}\\ %
        \scriptsize The Future Laboratory \\ Tsinghua University \\%
        \scriptsize College of Information Sciences and Technology \\ Penn State University %
\and Xinyang Li\thanks{e-mail: lixinyang22@mails.tsinghua.edu.cn}\\ %
     \scriptsize Academy of Arts \& Design \\ Tsinghua University %
\and Yuanxi Sun\thanks{e-mail: syxi0120@gmail.com}\\ %
     \scriptsize School of Computer and Cyber Sciences \\ Communication University of China %
\and Xinyi Fu\thanks{e-mail: fuxy@mail.tsinghua.edu.cn, Corresponding author}\\ %
     \scriptsize The Future Laboratory \\ Tsinghua University %
\and Christine Qiu\thanks{e-mail: qiu-yh18@tsinghua.org.cn}\\ %
     \scriptsize  School of Electrical Engineering and Computer Science \\ The KTH Royal Institute of Technology %
\and John~M. Carroll\thanks{e-mail: jmc56@psu.edu}\\ %
     \scriptsize College of Information Sciences and Technology \\ Penn State University
     }

\abstract{Understanding and recognizing emotions are important and challenging issues in the metaverse era. Understanding, identifying, and predicting fear, which is one of the fundamental human emotions, in virtual reality (VR) environments plays an essential role in immersive game development, scene development, and next-generation virtual human-computer interaction applications. In this article, we used VR horror games as a medium to analyze fear emotions by collecting multi-modal data (posture, audio, and physiological signals) from 23 players. We used an LSTM-based model to predict fear with accuracies of 65.31\% and 90.47\% under 6-level classification (no fear and five different levels of fear) and 2-level classification (no fear and fear), respectively. We constructed a multi-modal natural behavior dataset of immersive human fear responses (VRMN-bD) and compared it with existing relevant advanced datasets. The results show that our dataset has fewer limitations in terms of collection method, data scale and audience scope. We are unique and advanced in targeting multi-modal datasets of fear and behavior in VR stand-up interactive environments. Moreover, we discussed the implications of this work for communities and applications. The dataset and pre-trained model are available at \url{https://github.com/KindOPSTAR/VRMN-bD}.%
} % end of abstract

\CCScatlist{
  \CCScatTwelve{Human-centered computing}{Visu\-al\-iza\-tion}{Visu\-al\-iza\-tion techniques}{Treemaps};
  \CCScatTwelve{Human-centered computing}{Visu\-al\-iza\-tion}{Visualization design and evaluation methods}{}
}

\begin{document}


\firstsection{Introduction}

\maketitle

%% \section{Introduction} %for journal use above \firstsection{..} instead
Emotion is the most powerful motivational force in humans, and it is significantly correlated with perception, attention, memory and learning. Moreover, emotions with specific expressive abilities are a crucial human trait~\cite{dolan2002emotion}. With the rapid rise and intensive discussion of the metaverse concept, a virtual environment that blends the physical and digital, driven by the internet, web technologies, virtual reality (VR), augmented reality (AR) and mixed reality (MR)~\cite{lee2021all}, an increasing number of companies, developers and consumers are showing great expectations and enthusiasm~\cite{sparkes2021metaverse}. Among them, VR technology and consumer-grade VR devices serve as an important user interface to access the metaverse. Meanwhile, the act of human-machine interaction in VR has also inspired developers to imagine more possibilities for the future. The study of affective computing is not only an important research issue in the field of human-computer interaction~\cite{picard1999affective}, but also an essential research topic for the metaverse. The metaverse, as a post-real world~\cite{mystakidis2022metaverse}, possesses more plasticity and possibility than the real world. Furthermore, the metaverse is envisioned not only as a service concept but also as a future space with diverse socialities. Therefore, it is obvious that the metaverse should contain various characteristics and emotional expressions. Among all emotions, fear has a significant impact on human behavior, decision-making, mental health and social life as one of the most important basic human emotions~\cite{lang2019cognitive}. The study of how to induce, enhance~\cite{jicol2021effects}, diminish, calm~\cite{harris2002brief}, and overcome~\cite{freeman2018automated} fear in virtual environments can help people better deal with fearful feelings in the virtual world. 

At present, important issues on fear in the metaverse from include datasets, perceptions and recognition, simulation, and evaluation. Datasets, especially natural behavior data, are the basis for the computation of fear. A high-quality dataset is essential for studies on identification~\cite{shao2019objects365}, simulation~\cite{downs2022google}, design of stimuli~\cite{jemiolo2021emotion}, design of interactions~\cite{kirk2016data} and other machine learning tasks~\cite{sun2022importance}. Perception and recognition of emotions has been one of the most challenging issues within the domain of affective computing~\cite{wang2022systematic}. The multi-modal recognition algorithm for fear in virtual environments has great application potential. First, it has a positive impact on the development of VR, where more accurate recognition results may lead to an enhanced sense of realism and immersion in the metaverse. Second, such recognition of specific emotions can help improve human-centered and user-centered designs~\cite{la2020human}. Finally, this recognition method might provide a useful reference for identifying other emotions. Another crucial issue for the metaverse is evaluation. The evaluation of the fear response is decisive for the enhancement of relevant issues in the metaverse, such as user experience~\cite{shin2022actualization}, social systems~\cite{wang2022metasocieties}, usability~\cite{xi2022challenges}, ethics, morals and justice~\cite{bibri2022social}. In fact, the identification of negative emotions such as fear is more challenging than the identification of positive emotions because of self-concealment~\cite{larson2015self}. 

In this article, we addressed the following main research questions (RQs):
\begin{description}
   \item[RQ1] How can a multi-modal nonperforming fear dataset be built utilizing VR horror games?
   \item[RQ2] How can multi-modal data be used to recognize users’ fears?
\end{description}

To address these RQs, we proposed (1) a rigorous user experimental protocol design to induce participants’ fear utilizing VR horror games; (2) multi-modal data collection, processing, annotation, and fusion to build a nonperformance fear dataset; and (3) a prediction algorithm of fear using multi-modal data. In addition, we also provided insights into the emotional challenges of the metaverse. In summary, we made the following main contributions:
\begin{description}
   \item[(1)] A multi-modal natural behavior fear dataset.
   \item[(2)] A novel approach to fear recognition.
\end{description}

\section{Related Work}\label{sec:Related Work}
\subsection{Fear-arousing Stimulation}\label{subsec:Fear of emotions}

Scholars who support the classification of emotions generally agree that humans have more than a dozen basic emotions that contain physiological elements~\cite{solomonOpponentprocessTheoryAcquired1980,izard1991psychology}. Six basic emotions identified in Ekman’s facial-expression research~\cite{shiotaEkmanTheoryBasic2016} are commonly accepted: anger, disgust, fear, happiness, sadness and surprise, of which fear is a strongly unpleasant negative emotion. Fear is defined as a multidimensional response in which a person has an immediate emotional reaction and subsequent cognitive response to a perceptible threatening stimulus in the environment. Fear often arises from the presence or presumed presence of danger. In other words, it is an experience induced by a stimulus~\cite{lynchNothingFearAnalysis2015}.

People are susceptible to experiencing fear when they perceive significant and personally relevant dangers in real environments~\cite{lang1984cognition}. In addition, mediated environments can also evoke fear in people. Cantor suggests that the link between horror media and the fears of its viewers can be explained by the principle of stimulus generation~\cite{cantorFRIGHTREACTIONSMASS2008}. If real-world stimuli evoke emotions, then the same stimuli portrayed in media will evoke the same or less stressful experiences~\cite{harrisonTalesScreenEnduring1999}. The Diagnostic and Statistical Manual of Mental Disorders (DSM-IV)~\cite{spitzer1994dsm} framework classifies fear-provoking factors into five categories: animals (e.g., dogs or spiders); environment (e.g., fire or floods); blood/injections/injuries (e.g., wounds or needles); situational factors (e.g., height, confined spaces or more specific spaces such as a doctor's office); and other factors (objectively harmless but disturbing stimuli such as distorted faces and loud noises). These categories serve as references for our selection of horror games.

When a person's fear is aroused, the following physiological responses usually occur: 1) physiological changes (e.g., sweaty palms, increased heart rate or trembling), 2) changes in facial expressions (e.g., widened eyes, contracted facial muscles or increased tone), 3) changes in the way information is processed (e.g., more likely to receive suggested protective measures), and 4) the tendency to take specific actions (e.g., seeking cover, hiding or running away)~\cite{kivikangas2018emotion}. This immediate physiological response can be used as an objective measure of fear, and physiological signals collected in real time, such as heart rate~\cite{sartory1977investigation}, heart rate variability~\cite{wu2019amusement} and conductance~\cite{tabbert2006dissociation}, can all identify individuals' fear states. These references of responses in fear provide suggestions for the experimental design of this study, especially the choice of data modalities.

\subsection{VR Horror Games}\label{subsec:VR Horror Games}

The illusionary mechanisms provided by VR allow users to react realistically to VR scenes, effectively providing an immersive experience. The first is the place illusion (PI), also known as ‘being there’ or ‘presence’, which refers to a sense of being in a real place. The second is the plausibility illusion (PSI), an illusion that the events being portrayed are actually happening~\cite{slaterPlaceIllusionPlausibility2009}. VR makes users believe they are in the environment of the game (PI) experiencing the scene as it is happening (PSI)~\cite{linFearVirtualReality2017}. With advances in VR technology, this illusionary mechanism is achieved primarily through head-mounted displays combined with precise motion tracking systems, allowing the user to experience an interactive 3D virtual environment~\cite{davis2009avatars}. As a result, players playing horror games through VR experience a much stronger sense of fear and anxiety than those playing in video mode~\cite{pallaviciniEffectivenessVirtualReality2018}. In general, studies based on 3D environments can obtain better results than 2D stimuli~\cite{kakkos2019mental}. Further, a study by Somarathna et al.~\cite{9792298} showed that the fear could be effectively induced in VR environments, especially games.
    
Generally, horror games require players to actively respond to threats to survive. After a fear response is generated, people tend to use various coping measures to alleviate their fear. A survey by Lynch and Martins on video games~\cite{lynchNothingFearAnalysis2015} revealed that the stimuli that participants most often reported triggering their fears in games were darkness, disfigured humans, zombies and the unknown. Lin's proposed theoretical framework~\cite{linFearVirtualReality2017} for how players react to horror content in VR contains three strategies. 1) Approach (monitoring) strategies. At the cognitive level, players always alertly monitor their surroundings for possible hazards, while at the behavioral level, they choose to be proactive, for example, by adopting in-game skills to proactively eliminate hazards. 2) Self-help strategies. While on the cognitive level, players actively talk to themselves to encourage themselves, the behavioral level is where players choose to express their fears by screaming or shouting. 3) Avoidance strategies. The first of two avoidance strategies is physical and mental disengagement, where players usually turn their heads, close their eyes, remove their headphones or crouch down to avoid the sounds or images that frighten them, and the second is denial, where they tell themselves the experience is not real~\cite{linFearVirtualReality2017,zuckerman2003cope}. The findings in these past studies provided references for game selection and data annotation.

The exploration of horror content in VR can be applied in many ways. Virtual reality exposure therapy (VRET) is an increasingly common treatment for anxiety and specific phobias (agoraphobia, fear of driving, claustrophobia, aviophobia and arachnophobia)~\cite{parsonsAffectiveOutcomesVirtual2008} . Commercial games combining horror genres and biofeedback technology may be a useful stressor for practicing stress management skills~\cite{bouchard2012using, pallaviciniEffectivenessVirtualReality2018}. Additionally, experiencing scares in an entertainment environment has gained popularity by the market. Through an in-depth study of the characteristics of horror game content in VR games and what game elements induce fear in players and to what extent, this study provides a theoretical basis for fear computing research and a better product development evaluation tool for VR content designers and related researchers.

\subsection{Affective Computing Methods}\label{subsec:Affective Computing Methods}
\subsubsection{Body Gesture based Method}\label{subsubsec:Body Gesture based Method}

The possibility of emotion recognition through body gesture information has been widely acknowledged by researchers~\cite{vrigkas2015review}. Based on camera images, Cui et al.~\cite{keshari2019emotion} extracted the key points of human posture and used an algorithm to draw a simplified map to identify the emotional information for various postures. Shi et al.~\cite{shi2020skeleton} used the pose estimation algorithm to extract 3D skeleton information in the IEMOCAP database and proposed a self-attention enhanced spatial temporal graph convolutional network for emotion recognition of 3D skeletons, which confirmed the ability to recognize emotion based on 3D skeletons. Similarly, based on bone detection technology, Tsai et al. ~\cite{tsai2021spatial} trained the ST-GCN recognition model to effectively identify four emotional states. Sapiński et al.~\cite{sapinski2019emotion} proposed a method to recognize basic emotional states. It generates a model of affective action based on features inferred from the spacial location and the orientation of joints within the tracked skeleton. It shows the feasibility of automatic emotion recognition from sequences of body gestures, which can serve as an additional source of information in multi-modal emotion recognition. However, emotion recognition based on gestures is mostly used for specific single gestures, and its performance is greatly degraded for gestures in natural situations~\cite{sapinski2019emotion}. Bhattacharya et al.~\cite{bhattacharya2020step} presented a novel classifier network called STEP to classify perceived human emotion from gaits, and this model classified the perceived emotion of humans into one of four emotions: happy, sad, angry, or neutral. Fu et al.~\cite{fu2021gesture} presented an approach that used a Kinect v2 sensor to capture whole-body postures and recognize human fear in a VR environment using a long- and short-term memory model (LSTM). Previous studies have demonstrated that human emotions can be effectively recognized through skeletal poses. However, many researchers have claimed that current research is limited by the lack of video quality and have expressed concerns about the limitations of using a single modality. Therefore, combining information from other modalities and better datasets may be effective in improving performance.

\subsubsection{Audio based Method}\label{subsubsec:Audio based Method}
Speech is one of the most direct and oldest ways for humans to convey emotions, and voice information has been used by researchers for emotion research for more than 20 years~\cite{chen1998multimodal}. Additionally, voices displayed because of fear tend to be strong~\cite{tsuchiya2007emotion}, which seems to be a high-performance modality to apply to emotion recognition.
AnjaliBhavan et al.~\cite{bhavan2019bagged} proposed a bagged ensemble comprising support vector machines with a Gaussian kernel for SER. It extracted Mel-frequency cepstrum coefficients (MFCCs) and spectral centroids to represent emotional speech, followed by a wrapper-based feature selection method to retrieve the best feature set. Its best accuracy is 75.69\% using the RAVDESS database, which shows its superiority over other technologies such as AdaBoost. It should be noted that this work focuses on only acoustic features rather than speech features. Tzirakis et al.~\cite{tzirakis2018end} proposed a convolution recurrent neural network structure for speech emotion recognition. This model is made of a Convolutional Neural Network (CNN), which extracts features from the raw signal data and is stacked with a 2-layer LSTM. The model achieved the best results at the time regarding the consistency correlation coefficient for the RECOLA database. However, emotion recognition using voice alone is still insufficient because while voices may have distinctive features during strong emotions (crying, screaming, anger, etc.), it is often difficult to discriminate voices expressing nearly neutral emotions~\cite{clavel2008fear}. In addition, as humans become more socially engaged, people are increasingly suppressing the release of emotions through their voices~\cite{watkins1980silent}. Therefore, voices are worth using as valid information for identifying human emotions but should not be used as the only type of information.

\subsubsection{Physiological Signal based Method}\label{subsubsec:Physiological Signal based Method}
The application of physiological signals is considered an effective emotion recognition method. Many studies use physiological signals as the signal source for emotion recognition, but few studies use a physiological signal as the only criterion for emotion recognition. In previous studies, there have been two research methods for emotion recognition involving physiological signals. One approach is to use the combination of multiple physiological signals as the basis for emotion classification, such as the combination of respiration (RSP) and heart rate variation (HRV). Another approach is to combine physiological signals with signals from other modalities such as, facial expressions and voice. Oh et al.~\cite{oh2020design} used five physiological data, including respiration and heart rate, to optimize the model based on CNN to classify and identify 6 emotional states. Santamaria-Granados et al.\cite{santamaria2018using} applied a deep convolutional neural network on an AMIGOS dataset of physiological signals, which contains electrocardiograms and galvanic skin responses. It correlates physiological signals with the data of arousal and valence of the dataset and extracts the features of physiological signals in the domain of time, frequency, and nonlinearity. It made affective state prediction possible using physiological signals. The study by Moghimi et al.~\cite{8078217} demonstrated that physiological signals can be used effectively for emotion recognition classification tasks in a gaming environment.

\subsubsection{Multi-modal Method}\label{subsubsec:Multi-modal Method}
In recent years, this multi-modal analysis approach has received attention from researchers and is considered a possible future direction, which is noteworthy. There is evidence that multi-modal data fusion for emotion recognition is an important approach to address the limitations of each single-channel analysis mentioned above~\cite{gao2020survey}.  
Sebastian et al.~\cite{sebastian2019fusion} presented fusion techniques on deep learning models for improved emotion recognition in multi-modal scenarios. Intramodality dynamics for each emotion are captured in the neural network designed for the specific modality. 

Based on the existing behavioral emotion recognition, Matsuda et al.~\cite{matsuda2018emotour} combined behavioral features such as eye movements and head movements with audiovisual signals, collected human voice signals during travel, and classified three categories of emotions, positive (excited, happy/pleased and calm/relaxed), negative (sleepy/tired, bored/depressed, disappointed, dis-tressed/frustrated and afraid/alarmed), and neutral. Validation of bimodal data combining behavior and sound for emotion recognition in a real-world setting. Keshari et al.~\cite{keshari2019emotion} integrated facial expression and upper body gesture data to achieve higher emotion recognition accuracy than using single gesture data. In some special work environments, such as the emotion detection of driving states, some studies have used the combination of physiological signals and speech signals to extract and analyze data features for emotion classification in the corresponding work environment. Studies~\cite{ali2018emotion} have shown that multi-modal data, similar to the combination of speech and physiological signals, are more helpful to improve the recognition ability of emotion recognition systems than a single modality. In addition, Siddharthd et al.~\cite{8713896} showed that multi-modal fusion data has the advantage of the scalability and easy feature extraction.

\section{User Experiment Design}
To address our RQs, we designed a rigorous user experimental protocol to induce participants’ fear and gather participants’ data utilizing VR horror games. To address RQ1, we provide a complete and comprehensive experimental procedure, which also includes the data acquisition scheme. To address RQ2, we provide a LSTM-based predictive model.
\subsection{Experimental Situation}
\begin{figure}[!ht]
  \centering
  \includegraphics[width=\columnwidth]{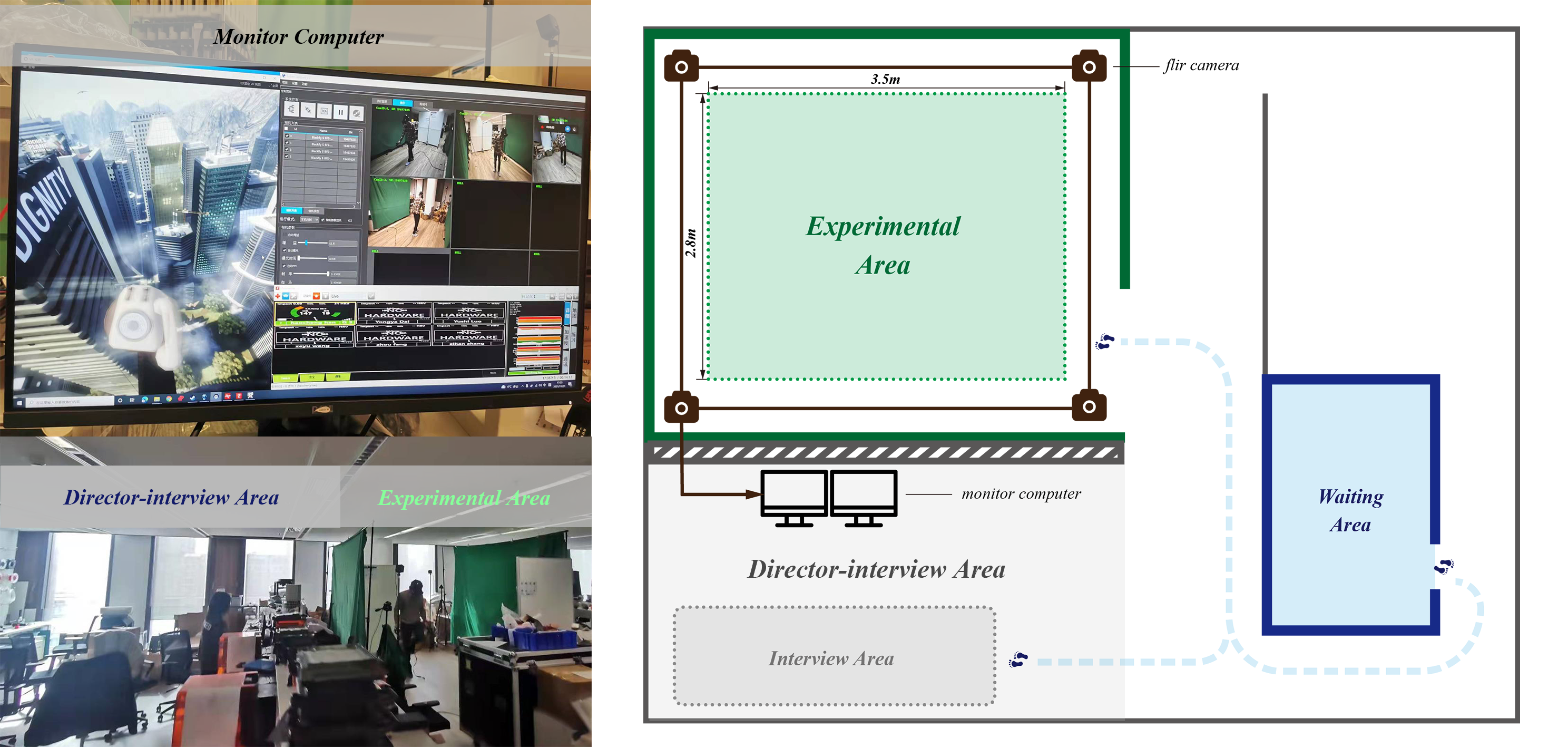}
  \caption{Experimental Places. The upper left corner shows the monitor computer screen, which included a real-time game screen, a multi-camera system monitoring screen, and a physiological signal recording equipment screen. The lower left corner shows the layout of the real experimental place, including the director-interview area and the experimental area. The right side shows a plan view of the full experimental place.}
  %\Description{Plane graph and real examples.}
  \label{fig.setup}
\end{figure}
The experimental site consisted of three areas: the waiting area, the experimental area, and the director-interview area (see Figure~\ref{fig.setup}). \textbf{Waiting area}: The waiting area was set up in a separate room, visually and aurally separated from the experimental and director-interview areas, where participants confirmed their attendance at the experiment, filled in the informed consent form and the pretest PANAS-X scale, wore the physiological signal acquisition equipment and waited for the experiment.
\textbf{Experimental area}: The experimental area was built in a 3.5m~$\times$~2.8m~$\times$~2.6m (Length~$\times$~Width~$\times$~Height) semiopen space partitioned by curtains and partition with the director-interview area. Experimental participants experienced the preinstalled VR games in this area. There were no obstacles in the space except for the VR equipment necessary for the experiment. There were four Filr cameras set up in each of the four corners of the experimental area, fixed at a certain height using a tripod. The VR device used in the experiment is the HTC Vive (with a per-eye resolution of 1080x1200, a refresh rate of 90 Hz, and a maximum field of view of about 110$^\circ$).
\textbf{Interview area}: The interview area was adjacent to the experimental area but isolated from it in terms of a realistic view; a set of PC computers and monitors were set up in this area to run the software required for various experiments, and the experimental host could view the situation in the experimental area in real time through the monitors. In addition, this area contained a subarea called the interview area for interviewing participants in various stages of the experiment.

\subsection{VR Horror Games Selection and Usage}
\begin{figure}[!ht]
  \centering
  \includegraphics[width=\columnwidth]{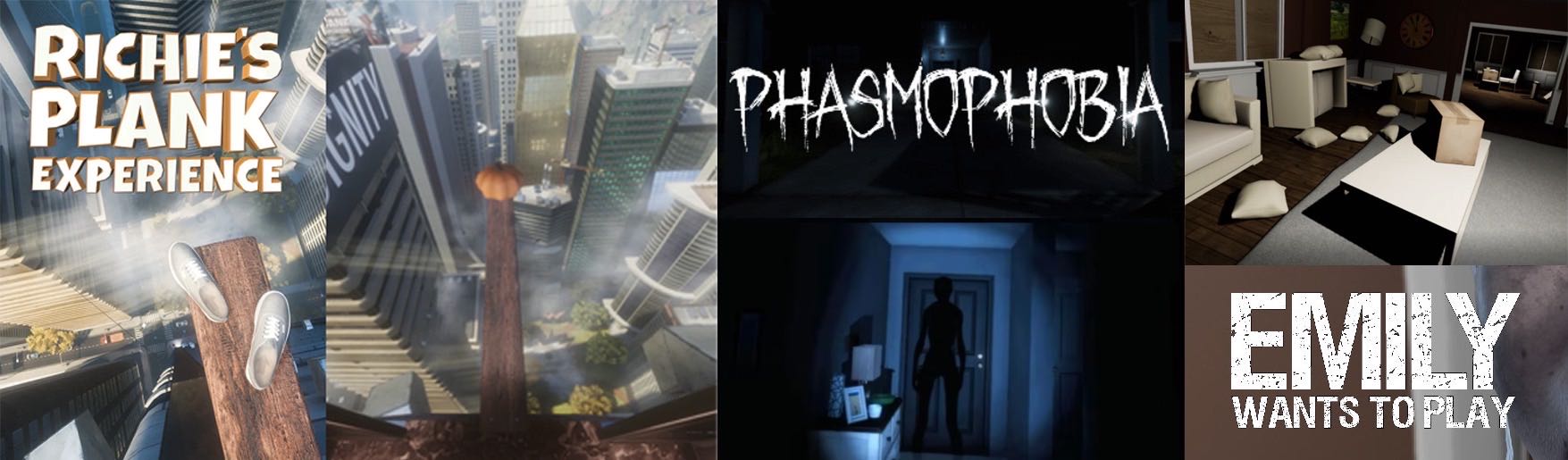}
  \caption{Examples of game posters and screenshots of the actual game. The first image on the left is the "Richie's Plank Experience" poster, the second image on the left is the "Richie's Plank Experience" in-game screenshot, the third image on the left is the "Phasmophobia" poster, the third image on the left is the "Phasmophobia" in-game screenshot, the first image on the right is the "Emily Wants To Play" in-game screenshot, and the first image on the right is the "Emily Wants To Play" poster.}
  %\Description{Real examples.}
  \label{fig.gamelist}
\end{figure}

In this experiment, the researchers chose three VR horror games from the Steam platform, Game 1 (\textit{Richie's Plank Experience}~\cite{toast2016richie}), Game 2 (\textit{Phasmophobia}~\cite{phasmophobia}), and Game 3 (\textit{Emily Wants To Play}~\cite{hitchcock2015emily}), which were thought to stimulate apparent fear. Figure~\ref{fig.gamelist} shows a schematic diagram of these games. The researchers chose to experiment with VR horror game rules as follows: 1) the game process is a linear script to control the game progress, length, and the same variables as much as possible; 2) neither short nor long-expected game time, each game processes within 5-20 minutes to avoid the failure or fading of emotional arousal caused by the length of the game; 3) the operation rules of the selected game are simple and can allow novices to learn quickly, and the game does not contain combinations of buttons and complex puzzles. 4) whether there are necessary elements to stimulate specific emotional fear; and 5) based on the player’s overall rating. 
Table~\ref{tab:gameselection} shows the metrics for each game based on the above criteria and whether the researcher provided spurious targets for controlling the flow of the experiment and triggering specific effects at the time of the experiment for reference. Considering the individual differences of the participants, to reduce the variables that were difficult to control in the game experiment, the researcher concealed or tampered with some real goals and provided a convincing false purpose to drive the participants to play the game~\cite{kors2015foundation}, which in turn made the plot progression of Game 2 and Game 3 independent of the player's behavior in the game. For example, in Game 2, players are asked to explore as many locations as possible to find a nonexistent red bear toy, while a brown bear can indeed be found in the game. This confusing task is set up to allow participants to explore as many dark environments as possible and trigger more potentially horrific game events. In Game 1, the player's actions based on the real game objectives did not branch the progress of the game, and therefore only the unique objectives were presented to the participants without changing the gameplay. %After Game 1, all participants were divided into two groups, with one group playing Game 2 before Game 3 and the other group playing Game 3 first, to avoid any effect of the order of playing of games containing similar potential fear triggers on the outcome.

\begin{table*}[!ht]
\centering
\caption{The selected games and their reference factors.}
\resizebox{\textwidth}{!}{%
\begin{tabular}{@{}ccccccc@{}}
\toprule
\begin{tabular}[c]{@{}c@{}}Game\\ No.\end{tabular} &
  Name of  game &
  \begin{tabular}[c]{@{}c@{}}Game (task) duration \\ (approximately)\end{tabular} &
  Ease of Learning &
  \begin{tabular}[c]{@{}c@{}}Types of Fears Simulated\\ (fear of)\end{tabular} &
  \begin{tabular}[c]{@{}c@{}}Player's overall rating\\  (positive rate - total user reviews)$^i$\end{tabular} &
  Spurious targets \\ \midrule
1 &
  \begin{tabular}[c]{@{}c@{}}Richie's Plank \\ Experience\end{tabular} &
  5 - 10 mins &
  Very Easy &
  \begin{tabular}[c]{@{}c@{}}Heights, Falling,\\ Spider, Dentists, \\ Jump scary\end{tabular} &
  \begin{tabular}[c]{@{}c@{}}Very Positive \\ (81\% - 567)\end{tabular} &
  No false goals \\
2 &
  Phasmophobia &
  20 mins &
  Easy &
  \begin{tabular}[c]{@{}c@{}}Claustrophobic/Darkness, \\ Solitude, Paranormal, \\ Death and Near-Death\end{tabular} &
  \begin{tabular}[c]{@{}c@{}}Overwhelmingly Positive \\ (96\% - 489,650)\end{tabular} &
  \begin{tabular}[c]{@{}c@{}}A false goal \\ and some false tips\end{tabular} \\
3 &
  Emily Wants To Play &
  5 - 10 mins &
  Very Easy &
  \begin{tabular}[c]{@{}c@{}}Claustrophobic/Darkness, \\ Paranormal, Jump scary, \\ Thunderstorm, Dolls\end{tabular} &
  \begin{tabular}[c]{@{}c@{}}Mostly Positive \\ (78\% - 1,694)\end{tabular} &
  \begin{tabular}[c]{@{}c@{}}A false goal \\ and some false tips\end{tabular} \\ \bottomrule
\multicolumn{7}{l}{$^i$Data collection date is September 18, 2023}
\end{tabular}%
}
\label{tab:gameselection}
\end{table*}

\section{Data Collection and Dataset Construction}

\subsection{Data Collection and Pre-processing}
\begin{figure}[ht]
  \centering
  \includegraphics[width=\linewidth]{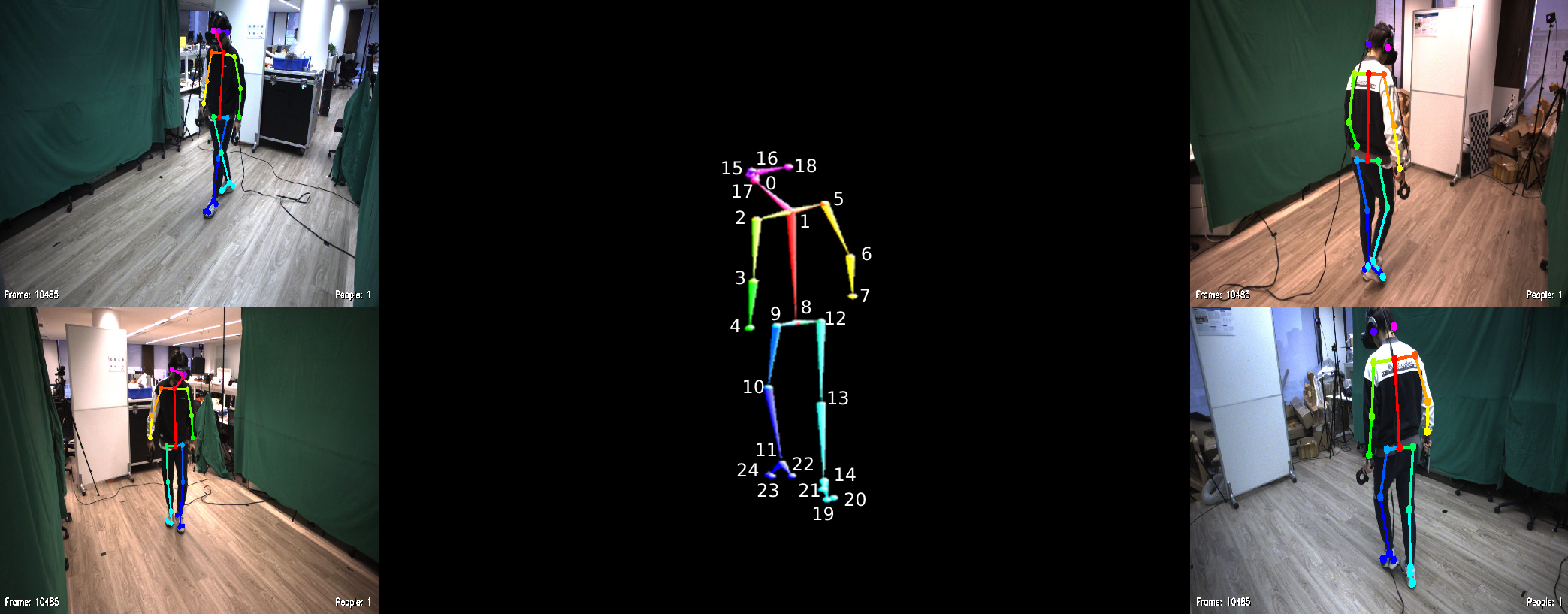}
  \caption{Human skeletal point calibration. Four Filr camera views (on two sides) and 3-D reconstructed skeletal point view (in the center). The alignment of each view and the reconstructed 3-D results were re-layout, but without modifying the content. The numbers in the mid figure represent the coordinates of the skeletal points, and the details are shown in the following Table~\ref{tab:openpose keypoints data}.}
  %\Description{Real examples.}
  \label{fig.Human skeletal}
\end{figure}

Our contributed dataset contains data from a total of three modalities: posture, audio and physiological signals. Four Filr cameras in the experimental area recorded the complete VR experimental process, with a single camera resolution of 1280 x 720. After calibrating the internal and external parameters of the camera, the human skeletal points were extracted using OpenPose~\cite{8765346} and reconstructed to obtain 3D keypoints. OpenPose takes RGB images as input in a single view camera and each key skeleton point in the image as output. OpenPose provides 25 key points, and Table~\ref{tab:openpose keypoints data} explains in detail the 25 key joints of the human body corresponding to the 25 key points. After 3D reconstruction of skeleton point data using a multiview camera system, 25 keypoints were represented by 3D coordinates (x-, y-, z-axis) with 75 factors in total. Missing values in 3D keypoints were further processed using interpolation. Figure~\ref{fig.Human skeletal} shows the views of each single camera in the experiment and the 3D pose estimation results. Participants wore wireless microphones on their lapels to record any vocals, and the audio was recorded at a bit rate of 128 kbps, covering the entire VR experiment process. Full game graphics and game sounds were recorded through the Xbox Game Bar. The physiological signal recording device (the Zephyr™ BioHarness™~\cite{hailstone2011reliability}) continuously recorded the heart rate and respiratory rate at a sampling rate of once every two seconds. The physiological data sampling rate provided by the device might not be perfect, but it is usable in this context~\cite{7868754,8882434}. Moreover, because this study employs multi-modal data, it reduces the limitations of relying on single modality data. At this stage, we filled in the missing values in the 3D skeletal points and aligned the data from different modalities according to the start and end timestamps, and split into three groups according to the different games.

% \usepackage{graphicx}
\begin{table}[ht]
\centering
\caption{Openpose (skeleton) keypoints data. Each key point corresponds to the example shown in Figure~\ref{fig.Human skeletal}. "R" means right, and "L" means left.}
\label{tab:openpose keypoints data}
\resizebox{\linewidth}{!}{%
\begin{tabular}{cccccccccc}
\bottomrule
\rowcolor[HTML]{EFEFEF} 
No.  & 0    & 1     & 2         & 3      & 4      & 5         & 6      & 7      & 8      \\
Name & Nose & Neck  & RShoulder & RElbow & RWrist & LShoulder & LElbow & LWrist & MidHip \\ \bottomrule
\rowcolor[HTML]{EFEFEF} 
No.  & 9    & 10    & 11        & 12     & 13     & 14        & 15     & 16     & 17     \\
Name & RHip & RKnee & RAnkle    & LHip   & LKnee  & LAnkle    & REye   & LEye   & REar   \\ \bottomrule
\rowcolor[HTML]{EFEFEF} 
No.  & 18   & 19      & 20        & 21    & 22      & 23        & 24    & \multicolumn{1}{l}{\cellcolor[HTML]{EFEFEF}} & \multicolumn{1}{l}{\cellcolor[HTML]{EFEFEF}} \\
Name & LEar & LBigToe & LSmallToe & LHeel & RBigToe & RSmallToe & RHeel & \multicolumn{1}{l}{}                         & \multicolumn{1}{l}{}                         \\ \bottomrule
\end{tabular}%
}
\end{table}

Then, we re-layout the recording of game footage with sounds, multicamera views, 3D keypoint reconstruction, and visualizations of physiological data, and merge them into one video aligned at the first frame. Meanwhile, we add a reference line based on video time to the visualizations of the physiological data for future annotation purposes.

\subsection{Data Report of Participant}\label{subsec:data report of participant}
We recruited those who were interested in participating in the VR experiment as participants through social media groups and nearby universities in China, but for health and safety reasons, the eligible candidates for this experiment were controlled to be between 18-30 years old and self-reported to be in good physical health, mentally fit, and free of diseases or medical histories such as heart disease, visual problems, and vertigo. All participants were paid a certain amount of cash as thanks. All participants were given the option to stay after the experimental period for a free experience with VR games for additional time (with no special restrictions) without interfering with the experiment or the typical research environment. This experiment was processed under approval of university IRB. Informed consent was obtained from all participants.

There were 23 participants (P1-P23, 9 males and 14 females), aged between 18 and 28 years old (median age = 21) and without significant congenital disorders. In total, 95\%(22) participants completed Game 1, 60\%(14) participants completed Game 2, and 56\%(13) of participants completed Game 3. The main reason for participants not completing the game was early withdrawal because of sensory overload. One person quit early due to intense discomfort caused by 3D vertigo/cybersickness, and we did not use data from this participant. The rest all reported early exit due to feeling too scary. In the cases of early withdrawal due to sensory overload, one person occurred during the game (shouted to stop at the very beginning of Game 2), and all the rest ended after the previous game interview (did not start the next game). %All participants completed the pretest and posttest Panas-X scales. Each participant who experienced a specific game completed the questionnaire for the corresponding game.

\subsection{Data Annotation}\label{subsec:dataannotation}
%\begin{figure}[htbp]
%  \centering
%  \includegraphics[width=1\linewidth]{graph/re-layout view-tool-sample.png}
%  \caption{Example of Re-layout Video and Annotation Tools. On the top left corner of the figure is the display area of the re-layout video, the area around the display area on the bottom right side is the annotation toolbar, and below the picture is the label recording area.}
%  %\Description{Real examples.}
%  \label{fig.Re-layout annotation tool}
%\end{figure}
We built an ancillary tool for helping manual annotation in the Windows platform. The tool allows annotators to annotate multi-modal data and improve the consistency and workability of the annotation process. The tool contained the area of monitoring, the annotation toolbar, and the labeled records. The data were annotated by watching video of participants' game views, multiview physical movements and the chart of physiological data combined with game sounds and microphone sounds by timestamp (in milliseconds). Annotators could use this annotation tool to replay video clips and change the annotation level (repetitive ratings). The annotated data included two categories (nonfear and fear), with 6 levels (this references the study by Fu et al.~\cite{fu2021gesture}, where level 0 $=$ nonfear, and levels 1-5, fear level from the lowest to highest). Among them, nonfear emotions (level 0) were automatically filled after annotation and no special manual annotation was needed. 

We recruited 5 annotators, and each reformatted video had at least 2 annotators. Before starting the annotation, a tutorial session and a simulated annotation using the sample were conducted for all the annotators to allow them to fully understand the usage of the annotation tool. We introduced some ground truths about body gestures~\cite{coulson2004attributing}, screaming~\cite{scheveneels2021predicting} and physiological signals~\cite{lonsdorf2017don} of fear 
conditioning to annotators during the simulation annotation. Also, the annotator considers the participant's self-report in the annotation process. After completing the annotation, we used absolute majority voting to obtain the final annotation results. If the absolute majority system failed, the final annotation results were equal to the nearest whole number of the average of all annotation levels (less than 1 to make up for 1). Finally, the annotation results were merged with the multi-modal dataset aligned by timestamp.

Here, we did not include participants' subjective fear ratings reported in real time in the dataset directly. Although self-reports have been used in numerous studies, there are some obvious limitations~\cite{vsalkevicius2019anxiety, bettis2022digital} that do not fit our dataset. Again, because collecting continuous user self-reports in VR environments, especially in motion, in a transient and precise manner poses a significant challenge in the form of user distraction~\cite{10.1145/3411764.3445487}.

\subsection{Data Analysis and Feature Extraction}

To exploit deep learning methods to predict fear levels, we structured these unstructured data through feature extraction for each modality. Then, we synchronized them according to the frame index, which was the simplest component of the video.

For the video model: we adopted OpenPose to learn 25 key human skeletal points in 3D spatial coordinates (d=75). Considering that these skeleton features were likely to contain a massive amount of redundant information, we therefore took the PCA approach to reduce the dimension from 75 to 33 while retaining 98$\%$ of the information as key features for each frame.

\begin{figure}[ht]
  \centering
  \includegraphics[width=0.95\linewidth]{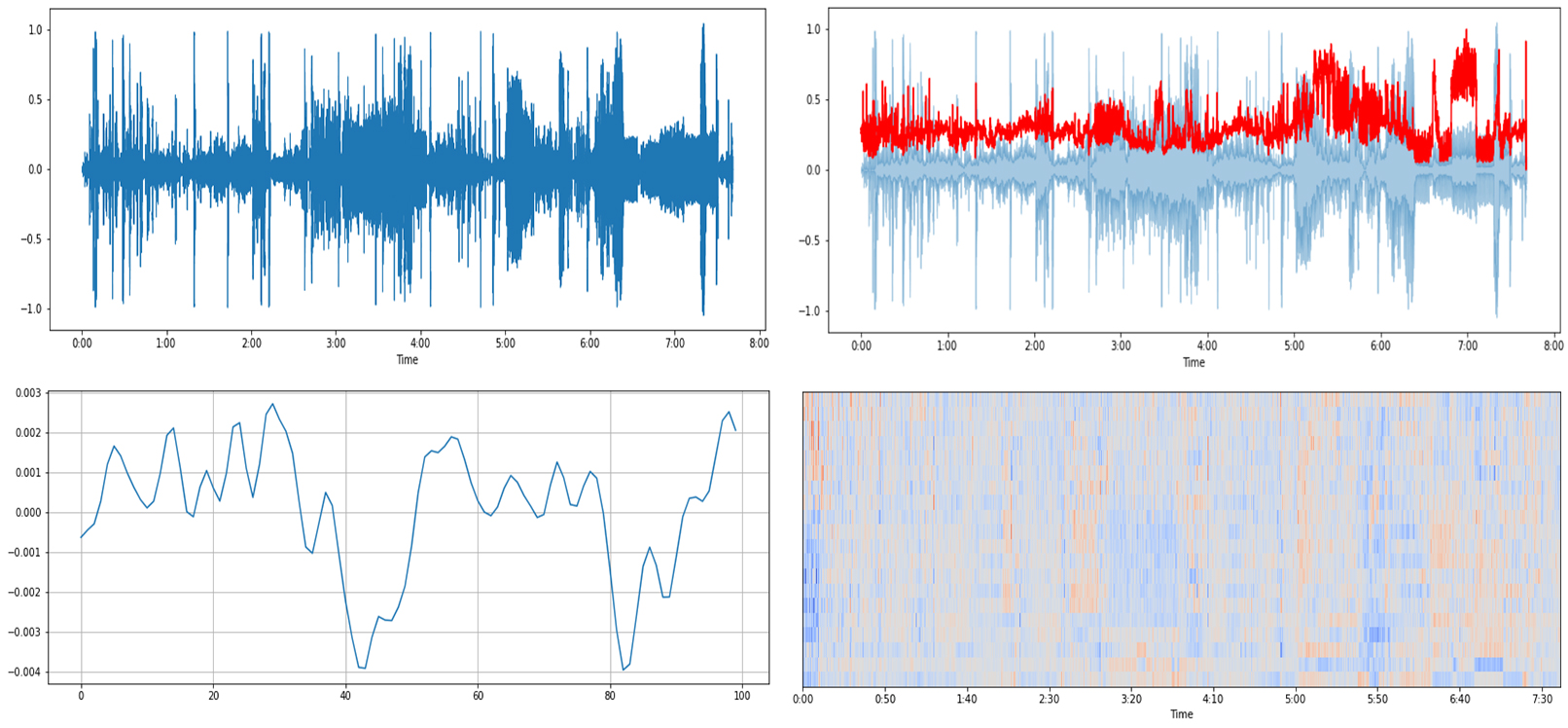}
  \caption{Example of Digital Audio Information, where the top left is the audio waveform plot, the top right is the spectral csentroid visualization, the bottom left is the zero-crossing rate (ZCR) visualization, and the bottom right is the Mel-frequency cepstral coefficients (MFCC).}
  %\Description{Real Audio examples.}
  \label{fig.Audio feature sample}
\end{figure}

For audio information: Game audio and microphone audio were extracted together and converted into a digital audio signal, and a sample of audio features is shown in Figure~\ref{fig.Audio feature sample}. The features used 7 metrics to represent the data: the zero-crossing rate, spectral centroid, spectral bandwidth, spectral rolloff, chroma features, rmse and MFCC. These metrics not only reflect the audio in terms of frequency, but also indicate human voices through the MFCC specifically. To align with skeletal features, we processed the audio features into framewise instead of secondwise features in 26 dimensions, of which 20 dimensions were MFCC features, and took the average of the audio features to align with the image frame rate. 

Similarly, we transformed the physiological data to describe the heart rate and breathing frequency of a participant per frame rather than per minute. The physiological data were aligned with the image frame rate using the average interpolation of adjacent data to obtain a complete dataset containing audio features and physiological data based on 3D keypoint data. Then, we concatenated the features in three modals as well as the target labels in accordance with the frame index to compose the complete feature data. 

\subsection{Dataset Construction}\label{subsec:Dataset Construction}
The video and audio durations of the dataset were 9 hours, 28 minutes and 58 seconds and contained 967079 frames in total. Our dataset consisted of 61 dimensions of 3-modal features and 3 dimensions of fear level labels. Among the features, 33 dimensions were extracted from 3D keypoints, 26 dimensions described audio characteristics, and 2 dimensions were related to physiological data. The labels contained the results of 2 annotators and their average score. After data annotation, we obtained a glimpse of the distribution of fear levels in the entire dataset. According to Table 4, participants barely showed any indications of being scared 58.18$\%$ of the time, which was labeled as level 0. The proportion of the remaining fear categories decreased with increasing level. Thus, the difficulty of the prediction task was increased to a certain extent due to the uneven distribution of the dataset. This also led to the attention mechanism being added to the model, which is discussed in the following section.
\begin{table}[ht]
%\vspace{-1.0em}
\centering
\caption{Description of our dataset. The first row indicates two different tasks (6- and 2-classification). The second row indicates emotion reference labels. In the third row, 0 to 5 are the fear levels(level 0 = non-fear; levels 1-5, fear level from the lowest to highest, or 1 = fear). The fourth row shows the number of fear annotated data in the dataset for each level. The fifth row represents the ratio of fear-annotated data of each level to the total dataset. Rows six to eight present the average heart rate and standard deviation for the different levels of fear annotated data. Row nine to eleven present the average respiratory rate and standard deviation for the different levels of fear annotated data. Rows twelve to fourteen introduce acceleration, referring to the average 3D skeletal point acceleration and its standard deviation, considering its movement in all three spatial dimensions (x-, y-, and z-axes). This acceleration is calculated from the individual accelerations in each of these dimensions. All values are rounded to two decimal places.}
\resizebox{\columnwidth}{!}{%
\begin{tabular}{c
>{}c 
>{}c 
>{}c 
>{}c 
>{}c 
>{}c 
>{}c 
>{\columncolor[HTML]{E8E8E8}}c 
>{\columncolor[HTML]{E8E8E8}}c 
>{\columncolor[HTML]{E8E8E8}}l 
>{\columncolor[HTML]{E8E8E8}}l 
>{\columncolor[HTML]{E8E8E8}}l 
>{\columncolor[HTML]{E8E8E8}}l 
>{\columncolor[HTML]{E8E8E8}}c }
\hline
 & \multicolumn{7}{c}{6-classification} & \multicolumn{7}{c}{\cellcolor[HTML]{C0C0C0}2-classification} \\ \hline
 & Non-fear & \multicolumn{5}{c}{Fear} &  & Non-fear & \multicolumn{5}{c}{\cellcolor[HTML]{E8E8E8}Fear} &  \\
 & 0 & 1 & 2 & 3 & 4 & 5 & Total & 0 & \multicolumn{5}{c}{\cellcolor[HTML]{E8E8E8}1} & Total \\
Count & 562681 & 284204 & 78099 & 31466 & 10202 & 427 & 967079 & 562681 & \multicolumn{5}{c}{\cellcolor[HTML]{E8E8E8}404398} & 967079 \\
Radio & 58.18\% & 29.39\% & 8.08\% & 3.25\% & 1.05\% & 0.04\% & 100\% & 58.18\% & \multicolumn{5}{c}{\cellcolor[HTML]{E8E8E8}41.82\%} & 100\% \\
\cellcolor[HTML]{C0C0C0}Heart rate &\cellcolor[HTML]{C0C0C0} &\cellcolor[HTML]{C0C0C0} &\cellcolor[HTML]{C0C0C0} &\cellcolor[HTML]{C0C0C0} &\cellcolor[HTML]{C0C0C0} &\cellcolor[HTML]{C0C0C0} &\cellcolor[HTML]{C0C0C0} &\cellcolor[HTML]{C0C0C0} &\cellcolor[HTML]{C0C0C0} &\cellcolor[HTML]{C0C0C0} &\cellcolor[HTML]{C0C0C0} &\cellcolor[HTML]{C0C0C0} & \cellcolor[HTML]{C0C0C0}&\cellcolor[HTML]{C0C0C0}\\
Mean & 94.39 & 97.42 & 97.26 & 98.09 & 104.30 & 92.62 & & 94.39 & \multicolumn{5}{c}{\cellcolor[HTML]{E8E8E8}97.61} & \\
Std & 17.11 & 17.50 & 17.92 & 16.15 & 21.24 & 4.80 & & 17.11 & \multicolumn{5}{c}{\cellcolor[HTML]{E8E8E8}17.61} &\\
\cellcolor[HTML]{C0C0C0}Breath rate &\cellcolor[HTML]{C0C0C0} &\cellcolor[HTML]{C0C0C0} &\cellcolor[HTML]{C0C0C0} &\cellcolor[HTML]{C0C0C0} &\cellcolor[HTML]{C0C0C0} &\cellcolor[HTML]{C0C0C0} &\cellcolor[HTML]{C0C0C0} &\cellcolor[HTML]{C0C0C0} &\cellcolor[HTML]{C0C0C0} &\cellcolor[HTML]{C0C0C0} &\cellcolor[HTML]{C0C0C0} &\cellcolor[HTML]{C0C0C0} & \cellcolor[HTML]{C0C0C0}&\cellcolor[HTML]{C0C0C0}\\
Mean & 15.89 & 16.71 & 17.82 & 17.91 & 18.38 & 16.86 & & 15.89 & \multicolumn{5}{c}{\cellcolor[HTML]{E8E8E8}17.06} & \\
Std & 5.61 & 5.24 & 5.77 & 5.74 & 5.73 & 5.31 & & 5.61 & \multicolumn{5}{c}{\cellcolor[HTML]{E8E8E8}5.42} &\\
\cellcolor[HTML]{C0C0C0}Acceleration &\cellcolor[HTML]{C0C0C0} &\cellcolor[HTML]{C0C0C0} &\cellcolor[HTML]{C0C0C0} &\cellcolor[HTML]{C0C0C0} &\cellcolor[HTML]{C0C0C0} &\cellcolor[HTML]{C0C0C0} &\cellcolor[HTML]{C0C0C0} &\cellcolor[HTML]{C0C0C0} &\cellcolor[HTML]{C0C0C0} &\cellcolor[HTML]{C0C0C0} &\cellcolor[HTML]{C0C0C0} &\cellcolor[HTML]{C0C0C0} & \cellcolor[HTML]{C0C0C0}&\cellcolor[HTML]{C0C0C0}\\
Mean & 0.28 & 0.09 & 0.09 & 0.09 & 0.09 & 0.12 & & 0.28 & \multicolumn{5}{c}{\cellcolor[HTML]{E8E8E8}0.09} & \\
Std & 51.08 & 0.28 & 0.57 & 0.07 & 0.55 & 0.12 & & 51.08 & \multicolumn{5}{c}{\cellcolor[HTML]{E8E8E8}0.35} &\\\bottomrule

\end{tabular}
}
\label{tab:description of our dataset}
\end{table}

\section{Multi-modal Fear Prediction Modeling}
To address RQ2, we adopted LSTM as our base model because LSTM can fully learn features in temporal dynamics and improve classification accuracy~\cite{8713896}. Building on this foundation, we extended the model by applying an additional backward layer of LSTM to enable learning in two directions. At the same time, the attention layer was introduced to generate the attention score through which frames that contain less useful information could be identified. Then, we showed the predicted results based on four different models according to our experimental data in 6- \& 2-classification tasks.

\subsection{Bidirectional LSTM + Attention Model}
The goal of our model was to predict fear levels for each frame based on a comprehensive integration of a sequence of multi-modal data. This includes skeletal points, audio features, and physiological data recorded in the same period of time. To this end, we adopted LSTM as our base model. On this basis, we extended the model by applying a backward layer of LSTM to learn in two directions. Thus, the model would be able to utilize information both from the past and future. In addition, the attention layer were also adopted to capture the most significant semantic information in the sequence. The attention mechanism operates by computing attention scores for each frame in the sequence.

\begin{figure}[H]
    \centering
    \includegraphics[width=\linewidth]{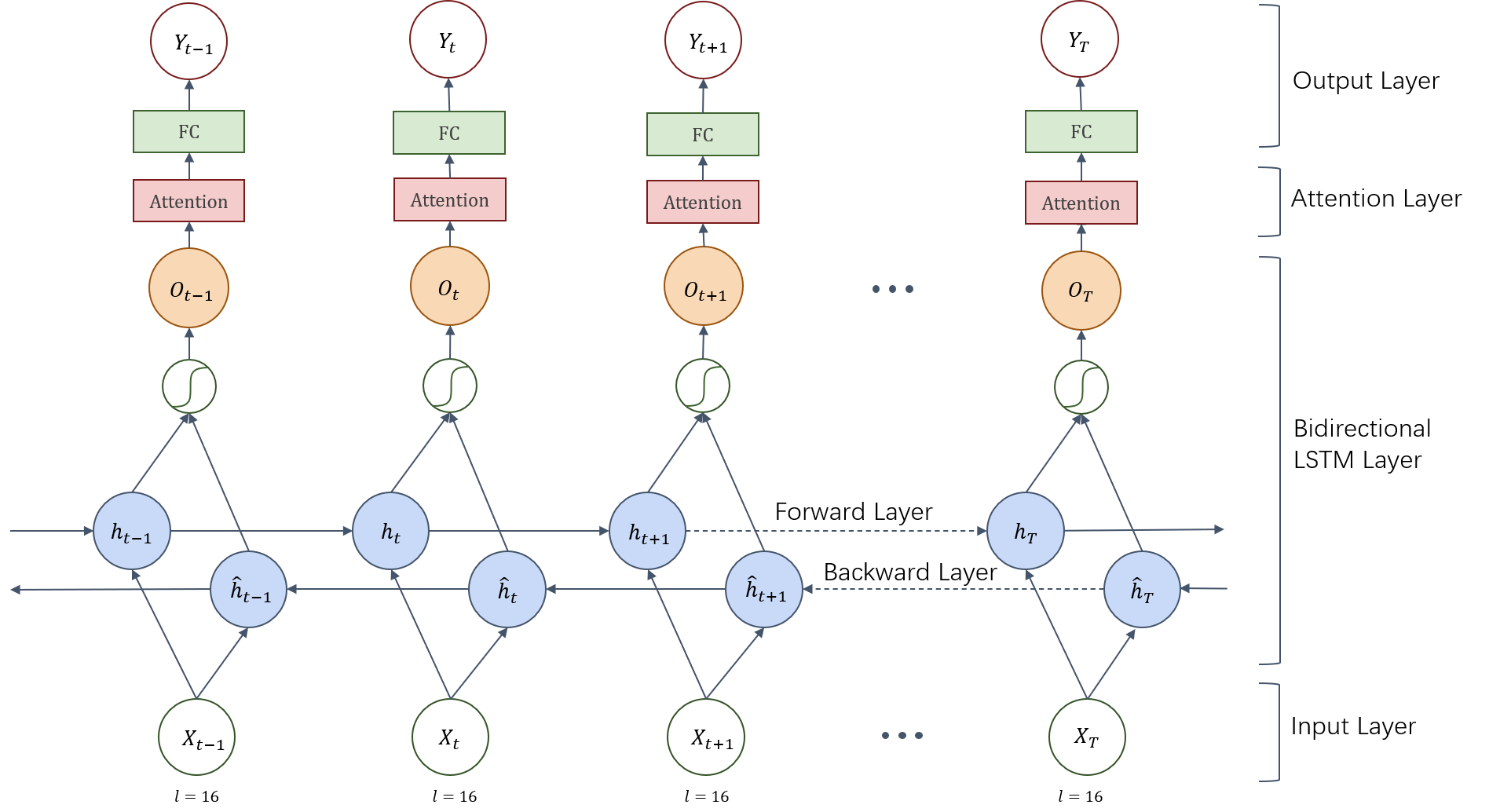}
    \caption{The architecture of BLSTM+attention model. $X_t$, $Y_t$ indicate the input and output on step $t$ of the model. $h_t$ and $\hat h_t$ stand for the hidden states of forward layer and backward layer for each step. $O_t$ is the corresponding output of BLSTM model.}
    \label{fig:bLSTM attention}
\end{figure}

As shown in Figure~\ref{fig:bLSTM attention}, the entire network consisted of 4 layers. In general, we now input a sequence of features  representing $l$ successive frames into the network and obtained the output as the fear level of the first frame in the sequence. These features encompass skeletal data, audio cues, and physiological responses, collectively providing a comprehensive dataset for analysis. The sequence could provide the information concerning the acceleration of human motions and the change of audio and physiological signals. Between the input and output layers, we established the BLSTM layer and the attention layer. In the BLSTM layer, there were two hidden states $h_t$ and $\hat h_t$ for both directions using the hidden state from the previous step and the input. This bi-directional processing is crucial for understanding the temporal context of fear responses, as it accounts for the progression and regression of emotional states. Following the BLSTM layer, the attention mechanism takes center stage. It computes attention scores for each frame by applying a learned transformation to the BLSTM outputs. The attention score calculation formula is:

\begin{equation}
\label{eqn:Intermediate-Representation-Computation}
    u_{t} = \tanh(state_t \times weight_W)~,
\end{equation}

where $u_{t}$ is the intermediate representation at time step $t$, $state_t$ is the output state of the LSTM at time step $t$, and $weight_w$ is the learned weight matrix for transforming the LSTM output state.

\begin{equation}
\label{eqn:Attention-Scores-Calculation}
    a^*_{t} = u_{t} \times weight_{proj}~,
\end{equation}

where $a^*_{t}$ represents the raw attention scores at time step $t$, $u_{t}$ is the intermediate representation computed as per Equation \ref{eqn:Intermediate-Representation-Computation}, and $weight_{proj}$ is another learned weight matrix used for projecting the intermediate representation onto the attention scores.

These scores are then normalized using the softmax function:

\begin{equation}
\label{eqn:spearman-brown1}
a_{ti} = \frac{exp(a^*_{ti})}{\sum_{i=0}^{l}exp(a^*_{ti})}~,
\end{equation}

where $a^*_{ti}$ represents the normalized attention score for the $i$th frame of sequence $t$. These scores signify the relative importance of each frame in the context of the entire sequence.

The model then computes a weighted sum of the BLSTM outputs, using the normalized attention scores as weights. This step effectively aggregates the sequence information, with a higher emphasis on frames deemed more relevant by the attention mechanism:

\begin{equation}
\label{eqn:spearman-brown2}
O^*_{t} = \sum_{i=0}^{l}a_{ti}O_{ti}~,
\end{equation}

where $O^*_{t}$ is the weighted output for sequence $t$. This attention-focused approach allows our model to be more sensitive and precise in predicting the fear level associated with each frame.

Finally, we utilized full connected (FC) layers to finish the classification job. Meanwhile, overfitting was a common issue in optimizing the model. To address the challenge of overfitting,we incorporated dropout as a regularization technique in the FC layers. Dropout randomly disables a fraction of neurons during the training process, which helps in preventing the model from becoming too dependent on specific features, thus enhancing its generalization capabilities.

\subsection{Prediction Results}
During the experiments, we randomly split the dataset into training (80$\%$), validation (10$\%$) and test (10$\%$) sets. The validation set was used to tune the hyperparameters, and we evaluated the model on the test set. To optimize the model, we tested models on different parameters.

We applied different combinations of these parameters to the LSTM, LSTM+attention, BLSTM and BLSTM + attention. Meanwhile, the performance was evaluated by 3 metrics: accuracy, recall and F1 score. According to Table~\ref{tab:comparison of approaches}, we found that the attention mechanism and the backforwad layer in BLSTM enhanced the predictive ability of the network to a certain extent for 6-classification task. 

In general, the BLSTM+attention reached the highest score on all three metrics for the 6-classification task, up to 65.31\% accuracy. The best result was obtained when the learning rate was 0.0001, the dropout rate was 0.5, the batch size was 256, the sequence length was 16, and we trained the model for 50 epochs. In addition, we perform a simple recoding of the dataset into a 2-classification task to test the performance of models in both fear and non-fear recognition tasks, and the accuracy is up to 90.47\%.

\begin{table}[htbp]
\centering
\caption{Comparison of the Approaches. The first row indicates two different tasks (6- and 2-classification). The second row indicates the model used. In this study, we focus on showing BLSTM+attention under 6-classification task. The third row indicates the accuracy of the model, where the accuracy of BLSTM+attention (6-classification) is 65.31\%. The fourth row indicates the recall of the model, where the recall of BLSTM+attention (6-classification) is 65.31\%. The fifth row indicates the F1 value of the model, where the F1 value of BLSTM+attention (6-classification) is 67.46\%. In addition, a reference to the results of the 2-classification task is provided on the right side of the table. All models were trained, tested and validated using the same dataset provided in this experiments.}
\resizebox{\columnwidth}{!}{%
\begin{tabular}{c
>{}c 
>{}c 
>{}c 
>{}c 
>{\columncolor[HTML]{E8E8E8}}c 
>{\columncolor[HTML]{E8E8E8}}c }
\hline
\multicolumn{1}{l}{} & \multicolumn{4}{c}{6-classification task} & \multicolumn{2}{c}{\cellcolor[HTML]{C0C0C0}2-classification task} \\ \hline
 & LSTM & LSTM+attention & BLSTM & BLSTM+attention & LSTM & BLSTM+attention \\
Accuracy & 60.22\% & 59.41\% & 61.90\% & 65.31\% & 90.47\% & 76.96\% \\
Recall & 59.69\% & 60.20\% & 61.74\% & 65.31\% & 90.47\% & 82.65\% \\
F1 & 61.34\% & 62.34\% & 63.96\% & 67.46\% & 90.47\% & 83.09\% \\ \hline
\end{tabular}%
}
\label{tab:comparison of approaches}
\end{table}

\section{Dataset Advancement}
We compared the dataset proposed in this paper with seven other datasets proposed in previous studies (see Table~\ref{tab:datasetcomparison}). Compared to previous studies, our dataset is unique, and has following advantages, because (1) our dataset contains more interactive behaviors in VR environments than watching videos~\cite{soleymani2011multimodal, xue2021ceap, yu2022eeg, tabbaa2021vreed, suhaimi2022dataset} or playing 2D games~\cite{kutt2022biraffe2}. (2) Only we provide the full body pose features. (3) We provide up to 6 categories of data for a specific emotion (fear). (4) We are the only dataset other than Soleymani et al.~\cite{soleymani2011multimodal} that provides audio features. (5) The largest data size in the VR game environment. (6) We are the only multi-modal database other than Granato et al.~\cite{granato2020empirical} that provides more interaction behaviors in VR environments. (7) We do not require participants to make self-reported annotations during the game, as a momentary and precise manner reporting would divide the user's attention in VR~\cite{10.1145/3411764.3445487}. However, we still provide self-reported data at the end of the game to assist data annotation.%supplement the dataset.

% \usepackage{graphicx}
\begin{table*}[!ht]
\centering
\caption{Comparison of datasets: The table shows seven representative datasets in the relevant fields from 2012 to 2022 and the basic information of our proposed dataset.}
\resizebox{\textwidth}{!}{%
\begin{tabular}{ccccccccc>{\columncolor[HTML]{E8E8E8}}c}
\hline
Author &  & Soleymani et al.~\cite{soleymani2011multimodal} & Xue et al.~\cite{xue2021ceap} & Granato et al.~\cite{granato2020empirical} & Yu et al.~\cite{yu2022eeg} & Tabbaa et al.~\cite{tabbaa2021vreed} & Kutt et al.~\cite{kutt2022biraffe2} & Suhaimi et al.~\cite{suhaimi2022dataset} & \cellcolor[HTML]{C0C0C0}\textbf{Ours (VRMN-bD)} \\ \hline
Year &  & 2012 & 2015 & 2020 & 2021 & 2021 & 2022 & 2022 & \textbf{2023} \\
Samples &  & 538 & \begin{tabular}[c]{@{}c@{}}32 participant \\ $*$ \\ 8 delected videos\end{tabular} & \begin{tabular}[c]{@{}c@{}}2 games \\ for each\\  participant\end{tabular} & \begin{tabular}[c]{@{}c@{}}120 trials $*$ 25 \\ participants $*$ 4 s\end{tabular} & 312 & $\approx$7650 mins & $\approx$20000 rows & \textbf{\begin{tabular}[c]{@{}c@{}}568 mins \\ and 58 seconds\\(967079 rows)\end{tabular}} \\
Data Fusion &  & Late Fusion & Pre-Fusion & Pre-Fusion & Single channel$^{\gamma}$ & Pre-Fusion & Dataset only$^{\gamma}$ & Pre-Fusion & \textbf{Pre-Fusion} \\
Subject$^{\alpha}$ &  & 27 & 32 & 33 & 25 & 26 & 102 & 32 & \textbf{23} \\
\begin{tabular}[c]{@{}c@{}}Stimulation \\ Method\end{tabular} &  & 2D Video & 360$^{\circ}$ VR Video & \begin{tabular}[c]{@{}c@{}}2D Game \\ and VR Game$^{\beta}$\end{tabular} & VR Video & 360$^{\circ}$ VR Video & \begin{tabular}[c]{@{}c@{}}2D Game, \\ Image and Audio\end{tabular} & 360$^{\circ}$ VR Video & \textbf{VR Game} \\
\begin{tabular}[c]{@{}c@{}}Active \\ Interactions\end{tabular} &  & $\times$ & $\times$ & $\checkmark$ & $\times$ & $\times$ & $\checkmark$ & $\times$ & \textbf{$\checkmark$} \\
\begin{tabular}[c]{@{}c@{}}Continuous\\ /Discrete\end{tabular} &  & C & C+D & C & C & C & C & C & \textbf{C+D} \\
\multirow{10}{*}{Modality} & Audio & $\checkmark$ & $\times$ & $\times$ & $\times$ & $\times$ & $\times$ & $\times$ & \textbf{$\checkmark$} \\
 & Body & $\times$ & $\times$ & $\times$ & $\times$ & $\times$ & $\times$ & $\times$ & \begin{tabular}[c]{@{}c@{}}\textbf{$\checkmark$} (33 features \\ * 3 dimensions$^{\delta}$)\end{tabular}   \\
 & Face & $\checkmark$ (20 features) & $\times$ & $\times$ & $\times$ & $\times$ & $\times$ & $\times$ & \textbf{$\times$} \\
 & Eye & $\checkmark$ & $\checkmark$ & $\times$ & $\times$ & $\checkmark$ & $\times$ & $\times$ & \textbf{$\times$} \\
 & EEG & $\checkmark$ & $\times$ & $\checkmark$ & $\checkmark$ & $\times$ & $\times$ & $\checkmark$ & \textbf{$\times$} \\
 & GSR & $\times$ & $\checkmark$ & $\checkmark$ & $\times$ & $\checkmark$ & $\checkmark$ & $\times$ & \textbf{$\times$} \\
 & HR\textbackslash{}HRV\textbackslash{}ECG & $\checkmark$ & $\checkmark$ & $\checkmark$ & $\times$ & $\checkmark$ & $\checkmark$ & $\times$ & \textbf{$\checkmark$} \\
 & BR & $\checkmark$ & $\times$ & $\checkmark$ & $\times$ & $\times$ & $\times$ & $\times$ & \textbf{$\checkmark$} \\
 & Acceleration & $\times$ & $\checkmark$ & $\times$ & $\times$ & $\times$ & $\checkmark$ & $\checkmark$ & \textbf{$\checkmark$} \\
 & Other & $-$ & SKT, BVP & EMG & $-$ & $-$ & $-$ & gyroscope & \textbf{\begin{tabular}[c]{@{}c@{}}screen \\ recording\end{tabular}} \\
\begin{tabular}[c]{@{}c@{}}Annotation \\ Method\end{tabular} &  & Self-report & Self-report & Self-report & Self-report & Self-report & Self-report & Pre-labeling & \textbf{Annotation} \\
\begin{tabular}[c]{@{}c@{}}Expression \\ Classification\end{tabular} &  & \begin{tabular}[c]{@{}c@{}}SBE plus Anxiety \\ and Amusement\end{tabular} & Valence-Arousal & Valence-Arousal & \begin{tabular}[c]{@{}c@{}}positive, neutral \\ and negative\end{tabular} & \begin{tabular}[c]{@{}c@{}}SEB plus Calm \\ and Anxious\end{tabular} & SEB plus Contempt & \begin{tabular}[c]{@{}c@{}}happy, scared, \\ calm, and bored\end{tabular} & \textbf{Fear} \\
\begin{tabular}[c]{@{}c@{}}Valence Level \\ of Classification$^{\epsilon}$\end{tabular} &  & 3+3 & 5 & 5 (in greneral) & 1 & 9 & 3+3+3 & 1 & \textbf{6} \\ \hline
\multicolumn{10}{l}{SBE = Seven Basic Emotions (anger, disgust, fear, happy, sad, surprise, and neutral).} \\
\multicolumn{10}{l}{$^{\alpha}$ The final number of participants} \\
\multicolumn{10}{l}{$^{\beta}$ Racing games: participants do not need to stand up and other actions.} \\
\multicolumn{10}{l}{$^{\gamma}$ Not applicable because of reasons.}\\
\multicolumn{10}{l}{$^{\delta}$ Contains 98\% of information.}\\
\multicolumn{10}{l}{$^{\epsilon}$ Maximum level.}
\end{tabular}%
}

\label{tab:datasetcomparison}
\end{table*}

In addition, we must acknowledge that the work has certain limitations, which are further discussed in section~\ref{sce:limitations}. However, we have four main aspects to support HCI and related communities~\cite{wobbrock2016research} thus far: (1) provides a specific approach to constructing multi-modal sentiment datasets. This paper presents a detailed construction process for a complete dataset, including experimental setup, data collection, annotation, analysis, and validation. By referring to the already validated construction process presented in this paper, future researchers can more easily construct multi-modal sentiment datasets in VR environments or in other environments. (2) provides an advanced multi-modal emotion dataset in VR fear game environments that contains continuous time series samples and rich features. Future researchers can use this dataset to test algorithm performance or as a reference for application development. (3) confirm the effectiveness of predicting fear emotion in virtual environments by combining body posture, audio, and physiological data. (4) Provides an annotation tool with a visual interface. This tool can effectively assist in the annotation of raw data, significantly reducing annotators' operational difficulty and annotation errors caused by asynchrony.

%\subsection{\rev{Dataset Sharing Statement}}
%\rev{The dataset, pre-trained model, demonstrations, and annotations are available at \url{https://github.com/KindOPSTAR/VRMN-bD}.}

\section{Discussions and Insight}
By reviewing past research and combining current trends in technology and society, we discovered that there is still a serious lack of discussion of emotional issues in the metaverse. VR, as one of the most likely metaverse building environments, was the focus of future research and development. Emotion was a part of effectively enhancing the realistic and social sense of the meta-universe. Hence, studies on user psychology and behavior in virtual scenarios are necessary. In this section, we first discuss the limitations of this paper. Then, we also discussed the challenges associated with emotions in the metaverse. 

\subsection{Limitations}\label{sce:limitations}
Our work has the following limitations. (1) We did not use EDA/GSR in our experiments, although it can be effective in identifying emotions~\cite{10.1145/3411764.3445370, lonsdorf2017don,10.1145/3534615}. The lack of mobility of GSR devices can create barriers to player interaction and degrade the gaming experience in VR gaming environments~\cite{dawson2017electrodermal}. Fortunately, we understand that there is much cutting-edge research on the use of gestures~\cite{10.1145/3411764.3445367,10.1145/3411764.3445442}, which makes it possible to abandon the use of joystick controllers in the future. Similarly, EEG signals are considered to be an effective channel for sensing human emotions, but based on the large conflict in wear compatibility between existing EEG devices and VR devices (especially VR head-mounted displays), using both devices at the same time can affect the quality of EEG signals due to player movements that cause the EEG to move or fall off and affect the VR gaming experience due to disruption of movement~\cite{xue2021ceap,10.1145/3281505.3281508}. Thus, although the above devices were not used in the construction of the dataset, a more realistic process of player experience was recorded. This dataset, because of its extremely high confidence level, could help in the development of wearable devices for virtual reality activities in the future. (2) We did not consider eyetracking in our experiments, although previous work has demonstrated a correlation between eye movements and emotion~\cite{geraets2021virtual}, and the potential for higher accuracy in multi-class emotion recognition tasks~\cite{soleymani2011multimodal, tabbaa2021vreed}. However, with the current technology, eyetracking VR devices are difficult to generalize in terms of cost~\cite{meissner2019combining, burch2022benefits}, which is beyond the current scope. We will investigate this work in the future. This problem will likely be solved as the usability and performance of wearable devices improve. However, wireless physiological measurements are still a recognized limitation~\cite{8432120} at this time. (3) Although a large number of strategies were used in this study to try to avoid individual differences, there is currently no effective way to completely eliminate individual differences. Among them, the algorithm is more difficult to recognize the low-fear annotation section. For this reason, the accuracy improved substantially after re-coding the multi-classification (6-classification) task into a 2-classification task. (4) We understand that running machine learning models consumes significant computational resources and that our proposed model and parameters may not be optimal solutions. Therefore, we plan to open-source the dataset for future research. In addition, we note the surprise/shock brought to the scientific community by the popularity of AI tools built on large language models (LLMs) since 2019, especially ChatGPT~\cite{radford2019language,brockman2016openai}. The tuning of models and their parameters for better performance through AI iterations has been achieved, and therefore the authors wish to emphasize the importance of datasets rather than the models themselves. (5) 3D vertigo is an important flaw in the VR experience~\cite{9757537} that can reduce the sense of presence in the VR experience~\cite{weech2019presence}. To best avoid the effect of 3D vertigo symptoms, we applied strategies to minimize the effects, such as selecting participants who self-reported "no" 3D vertigo for this study, and we did not find significant 3D vertigo symptoms for other participants during the experiment process. In addition, we have added intervals in different games to avoid 3D vertigo symptoms~\cite{ramaseri2022systematic}. We let participants play each game for no more than 20 minutes, and the shorter experience was an effective measure to avoid 3D vertigo symptoms~\cite{9854181}. However, it is still a challenge to avoid 3D vertigo symptoms especially for a long-term experience~\cite{tian2022review}. (6) The demographics of the participants may be a limitation. The participants were all from the Chinese population, which might introduce bias in the dataset, particularly due to different acquired fears and reactions in fear states caused by diverse cultural backgrounds and habits, and this could limit the ability to generalize the results to other groups of people. (7) The dataset proposed in this study is rich in information, yet this paper does not include all possible and interesting related research, especially those proven but in need of further exploration, such as studies on specific features (like postural acceleration) in relation to types of fear, research on the relationship between gender and levels of fear~\cite{zhang2023decoding}, and studies on the impact of social environment and psychological factors on behavior in playing VR horror games.

\subsection{Insights for Researchers and Game Developers}
In machine learning, the importance of high-quality datasets was obvious, especially in the context of the recent popularity of LLMs. The high-quality fear sentiment dataset constructed in this paper allowed developers to skip the complex and tedious data collection and thus study the specific problem directly (e.g., further modeling). The constructed model for identifying human fear emotions in VR environments would enable developers to verify the player's fear level in the game easily and further understand the player's fear response, which helped the design of game flow, hardware devices, security, etc. In real application development, researchers or developers can use our already trained models to understand players' fearful emotions and test whether games and applications successfully elicit fear from users. Furthermore, developers can use our research to design games and applications with different pacing, difficulty, or styles. We hope that the provided methods for collecting, processing, and applying time-series-based data, especially predictive models, offer the possibility to further consider the player experience to dynamically adjust the application scenario, difficulty and atmosphere. In addition, one of the purposes of our study using VR as a mediator is to consider the sense of realism and immersion. In other words, the player's response is closer to the actual scene. Therefore, the predictive models and theoretical contributions have a robust application in real life, such as creating an architectural atmosphere (e.g., haunted houses, escape rooms), therapy (overcoming fear), and understanding or scenario reenactment (real-life fears).

\section{Conclusion and Future Work}
Emotions have a significant impact on social life and human development. The development of VR and metaverse discussions have brought the topic of emotions in VR environments to unprecedented attention. People want to create a metaverse with more possibilities, and emotions in the virtual environment add more opportunities and humanity to this new "world". Emotions have a significant impact on social life and human development. The development of VR and metaverse discussions have brought unprecedented attention to the topic of emotions in VR environments. Our research addresses human fear emotions in virtual environments, providing an effective way to apply multi-modal data to identify fear emotions. To address these issues, we provide the complete experimental procedure, which includes game selection, experimental steps, data collection and labeling methods, dataset construction, and prediction model training. Overall, we provide a high-quality multi-modal (videos, audio, and physiological signals) immersive human fear responses dataset (VRMN-bD) and a fear prediction model with an accuracy of up to 65.31\% under 6-classifications, and accuracy of up to 90.47\% under 2-classifications task. We also provide a visual annotation tool for multi-modal data as a part of contribution to the research.

In addition to further optimizing the model and addressing the previously mentioned limitations, we hope to conduct future research on fear emotions and behavior strategies in other interesting scenarios. Moreover, we aim to extend this research to other types of emotions, and to compare and analyze fear and other emotional experiences in VR environments. This could help in identifying deeper relationships between various human emotions, especially in the context of the future metaverse. We plan to expand the number and diversity of participants in future studies, increasing the number of participants and considering unique horror elements in different cultural contexts, as well as the specific understanding and responses to fear among different groups of people. Finally, the dataset, pre-trained model, and more information are available at \url{https://github.com/KindOPSTAR/VRMN-bD}.

\acknowledgments{
This work was supported by the Youth Program of Chinese Ministry of Education Humanities and Social Sciences Project (Grant No. 23YJCZH049).}

\bibliographystyle{abbrv-doi}

\bibliography{template}

\end{document}